\documentclass[12pt]{article}
\usepackage{amssymb}
\usepackage{epsfig,color}
\usepackage{pstricks,graphicx,epsfig,color,amssymb,amsmath,amscd}
\usepackage{cite}
\usepackage[]{graphicx}
\usepackage{makeidx}
\usepackage{amsmath}
\usepackage{nccmath}
\usepackage{setspace}
\usepackage{ulem}
\usepackage{tensor}
\usepackage{bm}

\usepackage[]{caption}
\renewcommand\rho{\varrho}

\newcommand{\be}{\begin{eqnarray}}
\newcommand{\ee}{\end{eqnarray}}
\newcommand{\rar}{\rightarrow}

\begin{document}




\title{Superheavy SUSY-kind dark matter \\ and high energy cosmic rays}
\author{E.\,V. Arbuzova}
\date{}

\maketitle
\begin{center}

{\it arbuzova@uni-dubna.ru}\\

{Department of Higher Mathematics, Dubna State University,\\ 
Universitetskaya st.\,19, Dubna, 141983 Russia}\\
{Department of Physics, Novosibirsk State University,\\ 
Pirogova st.\,2, Novosibirsk, 630090 Russia}\\
\end{center}

\begin{abstract} 
The search for supersymmetric partners at Large Hadron Collider revealed negative result. Though, strictly speaking, it
does not exclude low energy supersymmetry, but still it leads to strong constraints of the parameter space. Therefore the search for 
supersymmetric particles at higher energies becomes of interest. It is shown that in $R^2$-modified cosmology heavy particles with the 
interaction strength typical for supersymmetry could be promising candidates for carriers of dark matter. 
We consider the heating of the Universe at the postinflationary stage via particle production by oscillating curvature scalar (scalaron). 
The bounds on the masses of dark matter particles are obtained for different dominant decay modes of the scalaron.
Possible impact of superheavy  particle decays on the spectrum of ultra high energy cosmic rays is discussed.
\end{abstract}

\maketitle

\section{Introduction}

The dark matter (DM) mystery is one of the most intriguing problems of modern cosmology. We know, that the dark matter is the invisible form of 
matter which discloses itself through its gravitational action. An accepted property of DM particles is that they are electrically neutral, since do not scatter light, so 
we call them {\it "dark matter"}, but other their properties are practically unknown (for review see 
\cite{Lin:2019uvt,Fox:2019bgz,Cline:2018fuq}). These facts open possibility for particles of many different types to be 
dark matter candidates. 

The value of the fractional mass density of dark matter according to observations \cite{Workman:2022ynf} is:
\be
\Omega_{DM} = \frac{\rho_{DM}}{\rho_{crit}} \approx 0.265.
\label{Om-DM}
\ee
 Here $\rho_{crit}$ is the critical energy density of the universe:
\be 
\rho_{crit} = \frac{3H_0^2 M_{Pl}^2}{8 \pi } \approx 5 \ {\rm keV/cm}^3, 
\label{rho-crit}
\ee 
where $M_{Pl}=1.22 \times 10^{19}$ GeV $ = 2.18 \times 10^{-5}$ g is the Planck mass\footnote{We use the natural system of 
units with $c=k=\hbar=1$.}  
and $H_0$ is the present day value of the Hubble parameter, for which we took
\be 
{H_0 = 100 h\, \rm{ km\, s}^{-1}\,\rm{Mpc}^{-1} \approx 70 \  \rm{km\, s}^{-1}\,\rm{Mpc}^{-1}.}
\label{H0}
\ee
Thus,  the observed dark matter energy density in contemporary universe is equal to: 
\be
\rho_{DM} = 1 \, {\rm keV/cm}^3.
\label{rho-DM-0}
 \ee 
  
 {The existence of dark matter and the magnitude of DM contribution into the total mass density 
 of the universe follow from the analysis of several independent pieces of data, which include:   }
\begin{itemize}
\item	flat rotational curves around galaxies;
\item equilibrium of hot gas in rich galactic clusters;
\item	 the spectrum of the angular fluctuations of Cosmic Microwave Background (CMB) Radiation;
\item onset of Large Scale Structure (LSS) formation at the redshift $z_{LSS}=10^4$ prior to hydrogen recombination at $z_{rec} = 1100$. 
\end{itemize}

Presently, possible carriers of dark matter are supposed to belong to two distinct groups: microscopically small (elementary particles) and macroscopically large.
The first group is abbreviated as WIMPs (Weakly Interacting Massive Particles) and contains axions with masses about $10^{-5}$ eV or even smaller, heavy neutral leptons 
with masses of several GeV, particles of mirror matter, the so-called superheavy  Wimpzillas, { the lightest supersymmetric
 particles,} and many others.  

The second group is presented by dark matter objects of stellar mass abbreviated as MACHO (Massive Astrophysical Compact Halo Object) and 
may include primordial black holes (PBH) with masses starting from $10^{20}$ g up to tenth solar masses, topological or non-topological solitons, and 
possible macroscopic objects consisting e.g. from the so-called mirror matter, {\it etc}.

Until resently, one of the most popular candidates for the role of DM carries was the Lightest Supersymmetric
 Particle (LSP) which, according to the low energy 
 {minimal supersymmetric (SUSY) model, should have mass of several hundred GeV, $M_{LSP} \sim $ 100 - 1000 GeV.  With the interaction strength typical for this}
 model the cosmological mass density of these particles was predicted to be close to the observed density of DM. However, 
 an extensive search for supersymmetry at Large Hadron Collider (LHC) led to negative results. The absence of 
 signal from supersymmetric partners at LHC, if not excluded, but considerably restricts the parameter space open for SUSY. 
 So if supersymmetry exists, its characteristic energy scale should be higher than, roughly speaking, 10~TeV.  
 
The cosmological energy density of LSPs is proportional to their mass squared: 
\be 
\rho_{LSP} \sim \rho_{DM}^{(obs)} (M_{LSP}/ 1\,{TeV})^2, 
\label{rho-LSP}
\ee
where $\rho_{DM}^{(obs)} \approx 1$ keV/cm$^3$ is the observed 
value of the cosmological density of DM,  see  Eq. (\ref{rho-DM-0}). 
For LSP with the mass $M_{LSP} \sim 1$ TeV, their energy density,  $\rho_{LSP}$,  would be
of the order of the observed dark matter energy density. For larger masses LSPs would overclose  the universe.
These unfortunate circumstances practically exclude LSPs as DM particles in the conventional cosmology. However, 
in $(R+R^2)$-gravity the energy density of LSPs may be much lower 
\cite{Arbuzova:2018ydn,Arbuzova:2018apk,Arbuzova:2020etv} and this fact reopens for LSPs the chance  to be 
the constituents of dark matter, if their mass  $M_{LSP} \geq 10^6$GeV 
(for  a review see \cite{Arbuzova:2021etq}).
 
\section{Cosmological evolution in $R^2$-gravity}

Theory of gravitational interactions, general relativity (GR), based on the Einstein-Hilbert action:
\be
{ S_{EH}= - \frac{M_{Pl}^2}{16\pi} \int d^4 x \sqrt{-g}\,R},
\label{S-GR}
\ee 
describes basic properties of the universe in very good agreement with observations. However, some features of the universe request to 
go beyond the frameworks of GR. In theories of modified gravity it is achieved by an addition of a non-linear function of curvature, $F(R)$, into 
the canonical action \eqref{S-GR}.  
 
In 1979  V.~T.~Gurovich and A.~A.~Starobinsky suggested to take $F(R)$ proportional to the curvature squared, $R^2$, for elimination of the 
cosmological singularity \cite{Gurovich:1979xg}. In the subsequent paper by Starobinsky  \cite{Starobinsky:1980te} it was found that 
$R^2$-term leads to exponential cosmological expansion  (Starobinsky inflation). The corresponding action has the form:
\be
S_{tot} = -\frac{M_{Pl}^2}{16\pi} \int d^4 x \sqrt{-g}\,\left[R- \frac{R^2}{6M_R^2}\right] + S_{matt},
\label{action-R2}
\ee
where $M_R$ is a constant parameter with dimension of mass and $S_m$ is the matter action. According to the estimate of 
Ref.~\cite{Faulkner:2006ub} the magnitude of temperature fluctuations of CMB demands
$M_R \approx 3\times 10^{13}$~GeV. In our paper \cite{Arbuzova:2018ydn} it was found that $R^2$-term 
creates considerable deviation from the Friedmann cosmology in the post-inflationary epoch and, thereby,  
kinetics of massive species in cosmic plasma and the density of DM particles differ significantly from those in the conventional cosmology.

The modified Einstein equations obtained from action \eqref{action-R2}  have the form:
\be  R_{\mu\nu} - \frac{1}{2}g_{\mu\nu} R -
 \frac{1}{3M_R^2}\left(R_{\mu\nu}-\frac{1}{4}R g_{v}+g_{\mu\nu} D^2-  D_\mu D_\nu\right)R
 =\frac{8\pi}{M_{Pl}^2}T_{\mu\nu}\,, 
 \label{field_eqs}
\ee
where $D_\mu$ is the covariant derivative, $D^2\equiv g^{\mu\nu} D_\mu D_\nu$ is the covariant D'Alembert operator, and
$T_{\mu\nu}$ is the energy-momentum tensor of matter.  

Taking trace of Eq.~(\ref{field_eqs}) yields
\be
D^2 R + M_R^2 R = - \frac{8 \pi M_R^2}{M_{Pl}^2} \, T^\mu_\mu.
\label{D2-R}
\ee
The general relativity limit can be recovered when $M_R\rightarrow \infty$.
In this case, we expect to obtain the usual algebraic relation between the curvature scalar and the trace of the energy-momentum tensor of matter:
\be
M_{Pl}^2 R_{GR} = - 8\pi T_\mu^\mu\,.
\label{G-limit}
\ee

We assume that the cosmological background is described by the spatially flat 
Friedmann-Lema\^itre-Robertson-Walker 
(FLRW) metric:
\be 
ds^2 = dt^2 - a^2(t) \delta_{ij} dx^i dx^j,
\label{FLRW}
\ee 
where $a(t)$ is the cosmological scale factor and 
\be
H = \frac{\dot a}{a} 
\label{h-of-a}
\ee
is the Hubble parameter at an arbitrary time moment. 

We consider the homogeneous and isotropic  distribution of matter with the linear equation of state: 
\be
P = w \rho,
\label{eq-state}
\ee
where $\rho$ is the energy density, $P$ is the pressure of matter,
and $w$ is usually a constant parameter. For non-relativistic matter 
$w=0$, for relativistic matter $w=1/3$, and for the vacuum-like state $w=-1$. 

Correspondingly, the energy-momentum tensor of matter $T^\mu_ \nu$ has the following  diagonal form:
\be
T^\mu_\nu = diag(\rho, -P, -P, -P).
\label{T-mn}
\ee

For homogeneous field, ${R=R(t)}$, and with equation of state \eqref{eq-state} the evolution of curvature is governed by the 
equation:
\be 
\ddot R + { 3H\dot R} +M_R^2R = - \frac{8\pi M_R^2}{M_{Pl}^2}(1 - 3w)\rho. 
\label{ddot-R}
\ee

In metric (\ref{FLRW}) the curvature scalar is expressed through the Hubble parameter as:
\be
R=-6\dot H-12H^2\,.
\label{R-of-H}
\ee

The energy-momentum tensor satisfies the covariant conservation condition, $D_\mu T^\mu_\nu = 0$, which in 
FLRW-metric (\ref{FLRW})  has the form: 
\be
\dot\rho = -3H(\rho+P)  = -3H (1+w) \rho\,.
\label{dot-rho}
\ee

Equation \eqref{ddot-R} for the curvature evolution does not include the effects of particle production by the curvature 
scalar. It  is a good approximation at inflationary epoch, when particle production by $R(t)$ is practically
absent, because $R$ is large  by the absolute value
and  the Hubble
friction is large, so  $R$ smoothly evolves down to zero. At some stage,
when $ H$ becomes smaller than $ M_R$, curvature starts to oscillate  around zero
efficiently producing particles. 
It commemorates the end of inflation and the onset of the heating of the universe, 
which was originally void of matter.
At that moment the transition from the accelerated expansion (inflation) to a de-accelerated one took place.
 The universe evolution at this stage is similar to
Friedmann matter dominated regime but still differs  from it in many essential
features.

 As we see in what follows, curvature, $R(t),$ can be considered as an effective scalar field, named {\it scalaron}, with the 
mass equal to $ M_R$, since the left hand side (l.h.s.) of  
equation of motion of $R(t)$ (\ref{D2-R})
exactly coincides with the l.h.s. of the Klein-Gordon equation
for  massive scalar particle.

It is convenient to introduce dimensionless time variable and dimensionless functions:
\be
 \tau =  M_R\,t,\ \ \ H = M_R\, h, \ \ \ R = M_R^2\, r, \ \ \ \rho = M_R^4\, y. \ \ \
\label{dim-less}
\ee
Equations (\ref{ddot-R}), (\ref{R-of-H}),  and (\ref{dot-rho}) now become:
\be
&&r'' + 3h  r' + r = - 8 \pi \mu^2 (1-3w) y, \label{r-two-prime}\\
&&h' + 2h^2 = - r/6, \label{h-prime} \\
&&y' + 3(1+w)h\,y = 0, \label{y-prime}
\ee
where prime means derivative over $\tau$ and $\mu = M_R/M_{Pl}$.

Let us consider first the inflationary stage of "empty" universe with $\rho =0$ ($y=0$ in dimensionless quantities).  
It is shown in Ref.~\cite{Starobinsky:1980te} that for sufficiently large $R$ the {devoid} of matter universe expanded quasi exponentially
long enough to provide solution of flatness, horizon and homogeneity problems existing in Friedmann cosmology (for the review see e.g. the book \cite{Linde:2005ht}).
 
According to definition \eqref{h-of-a} the scale factor is expressed through $h(\tau)$ as
\be
a(\tau) \sim \exp \left[  { \int_0^{\tau} h(\tau')\,d\tau'} \right].   
\label{a-inf}
\ee
If the Hubble parameter slowly changes with time, the cosmological expansion would be close to the exponential one and the scale factor 
during inflation would increase as:
\be 
\frac{a_{inf}}{a_0} = \exp \left[  { \int_0^{\tau_{inf}} h(\tau)\,d\tau}\right] = \exp [N_e]\,,
\ee
where $\tau_{inf}$ is the duration of inflation and $N_e$ is the so-called  number of e-foldings. The initial conditions should be chosen in such a way that 
at least 70 e-foldings are ensured:
\be
N_e = \int_0^{\tau_{inf}} h\,d\tau \geq 70.
\label{h-dt}
\ee
This can be  realised if the initial value of $r$
is sufficiently large  and practically independent on the initial value of $h$. 

We can roughly estimate the duration of inflation neglecting higher derivatives in 
Eqs.~(\ref{r-two-prime}) and taking $y=0$, so we arrive to the  following simplified set of equations:
\be
h^2 &=& - r/12, \label{h2}\\
3 h r'  &=& - r  . \label{h-r-prime}
\ee
These equations are solved as:
\be
\sqrt{-r(\tau)} =  \sqrt{-r_0} - \tau/\sqrt 3,
\label{sqrt-r}
\ee
where $r_0$ is the initial value of $r$ at $\tau = 0$. According to Eq. (\ref{h2}), the Hubble parameter  
behaves as $h(\tau)=(\sqrt{-3r_0} - \tau)/6$. The duration of inflation is roughly determined by the condition 
$h=0$:
 \be
 \tau_{inf} = \sqrt{-3r_0}.
 \label{tau-inf}
 \ee 

\begin{figure}[!htbp]
  \centering
  \begin{minipage}[b]{0.4\textwidth}
    \includegraphics[width=\textwidth]{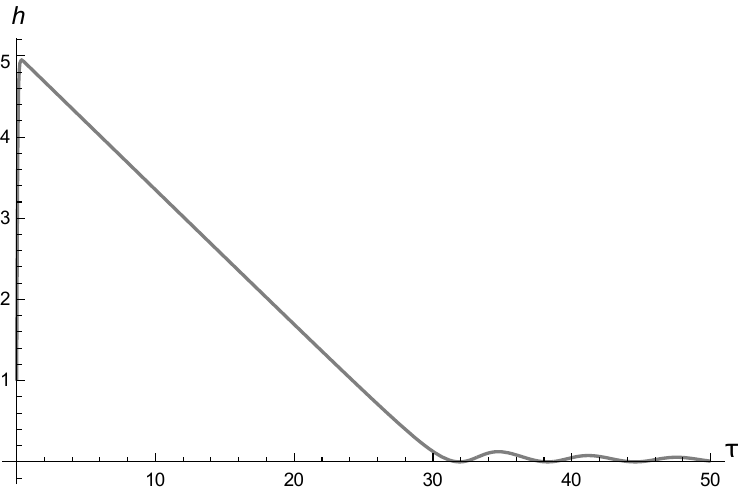}    
  \end{minipage}
  \hspace*{.15cm}
  \begin{minipage}[b]{0.4\textwidth}
    \includegraphics[width=\textwidth]{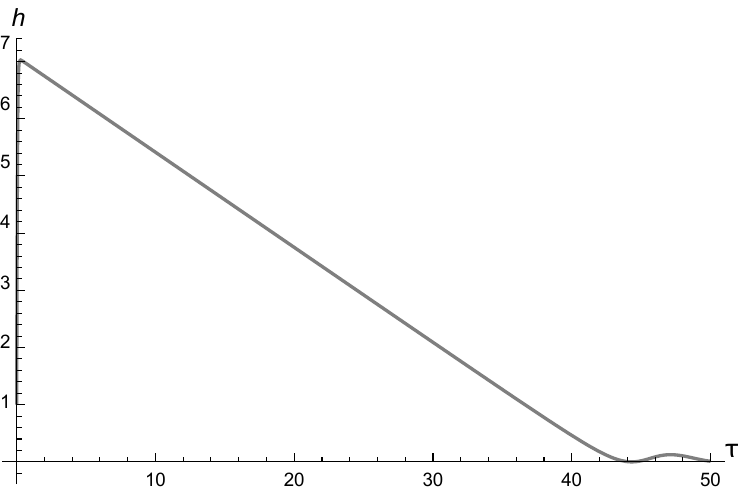}
      \end{minipage}
  \caption{Evolution of $h (\tau)$ at the inflationary stage with the initial values of dimensionless curvature
  $|r_0| = 300$ (left) and $|r_0| = 600$ (right). 
   The numbers of e-foldings, according to  Eq. (\ref{h-dt}), are respectively 75 and 150.}
  \label{f:h-dt-inf}
 \end{figure}

The number of e-folding is equal to the area of the triangle below 
the line $h(\tau)$, thus $N_e \approx |r_0| / 4$. It is in excellent agreement with numerical solutions of
Eqs.~(\ref{r-two-prime})-\eqref{h-prime} depicted in Fig.~\ref{f:h-dt-inf}.
After  the Hubble parameter, $H(t),$ reached zero, it started to oscillate around it with the amplitude decreasing as $2/(3t)$ and 
the exponential rise of a scale factor, $a(t)$, turns into a power law one.  

\begin{figure}[!htbp]
  \centering
    \includegraphics[width=0.45\textwidth]{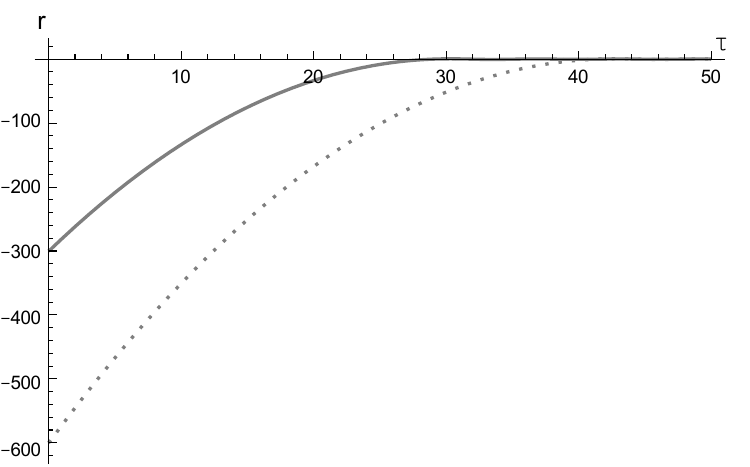}
    \begin{minipage}[b]{0.45\textwidth}
    \includegraphics[width=\textwidth]{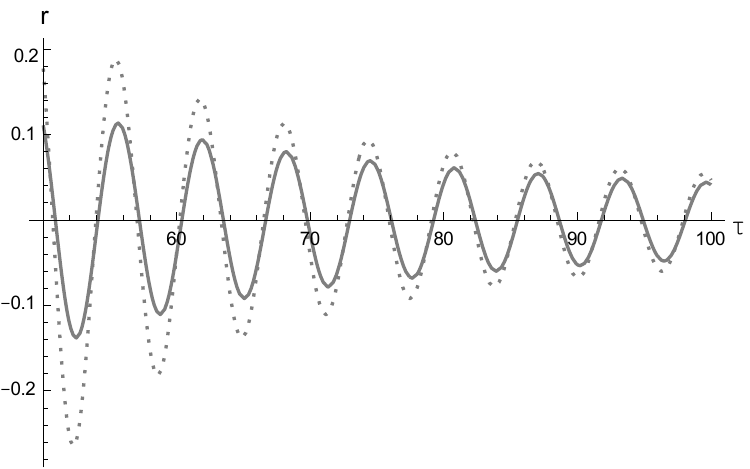}
\end{minipage}
 \caption{Evolution of the dimensionless curvature scalar for $r_0=-300$ (solid)
and  $r_0=-600$ (dotted). 
Left panel: evolution during inflation; right panel: evolution after the end of inflation when curvature scalar starts to oscillate
(scale differs from the left graph).
}
  \label{f:r36}
 \end{figure}

\begin{figure}[!htbp]
  \centering
  \begin{minipage}[b]{0.45\textwidth}
    \includegraphics[width=\textwidth]{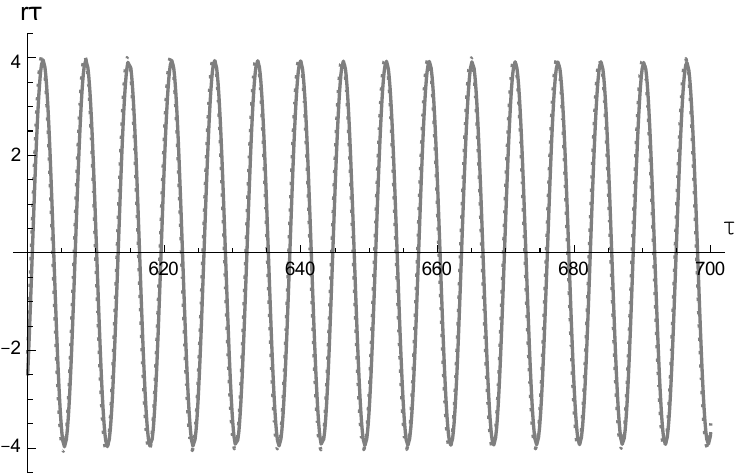}
      \end{minipage}
  \begin{minipage}[b]{0.45\textwidth}
    \includegraphics[width=\textwidth]{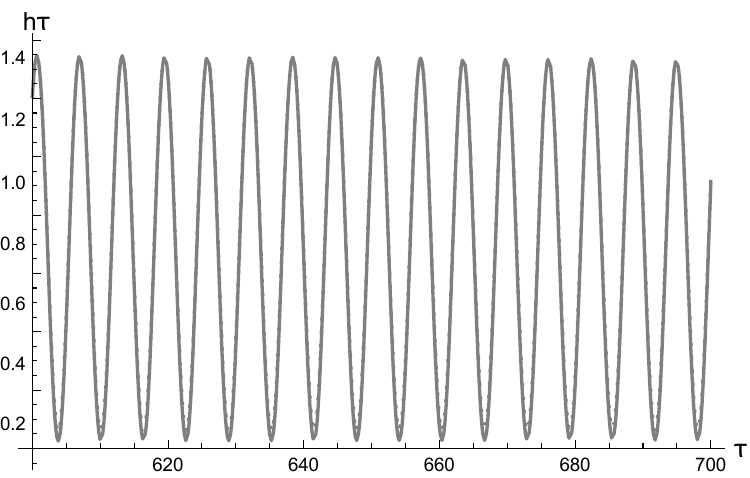}
\end{minipage}
  \caption{Evolution of the curvature scalar $ r(\tau)  \tau$ (left panel) and the Hubble parameter $ h(\tau)  \tau$ (right panel) 
  at post-inflationary epoch as functions of dimensionless time $\tau $. 
 }
  \label{f:r-postinf}
 \end{figure}

The evolution of the dimensionless curvature scalar, $r$, during inflation is presented in the left panel of Fig.~\ref{f:r36}. 
Inflation terminates when both, $h$ and $r$, reach zero. Numerical solution for $r(\tau)$ immediately  after the end of inflation  is presented in the right panel of Fig.~\ref{f:r36}. 
For larger $\tau $ the solutions, $r(\tau)\tau $ and $h(\tau)\tau $, take very 
simple forms depicted in Fig.~\ref{f:r-postinf}. Both $ r(\tau) \tau $ and $h(\tau)\tau $ oscillate with constant amplitudes. 
The curvature, $\tau r(\tau)$, oscillates 
around zero, while the Hubble parameter, $\tau h(\tau )$, oscillates around $2/3$ 
almost touching zero at the minima.

Stimulated by the numerical solutions 
we find  the asymptotic analytical solutions (for details see Ref.~\cite{Arbuzova:2018ydn}): 
\be
&&{r=-\frac{4\cos(\tau+ \theta)}{\tau} - \frac{4}{\tau^2}}, \label{rsol} \\
&&h= \frac{2}{3\tau} \left[1+ \sin(\tau + \theta )\right] \label{hsol}, 
\ee
where the constant phase $\theta $ is determined {from}
the initial conditions and 
can be adjusted by the best fit of the asymptotic solution to the numerical one.
Comparison of numerical calculations presented  in Fig.~\ref{f:r-postinf}   
with analytical estimates \eqref{rsol} and \eqref{hsol}
gives ${\theta = -2.9\pi/4}$.

Curvature oscillations give rise to  production of elementary particles. 
The back reaction of particle production on the evolution of curvature, $R(t)$, is usually described by an addition of 
the friction term, $\Gamma \dot R$, into the l.h.s. of Eq. (\ref{ddot-R}) 
with a constant~$\Gamma $:
\be 
\ddot R + (3H + { \Gamma}) \dot R+M_R^2R = - \frac{8\pi M_R^2}{M_{Pl}^2}(1 - 3w)\rho.
\label{ddot-R-Gam}
\ee
Here $\Gamma $ is the decay width of the scalaron. 
The value of $\Gamma $ depends on the concrete decay channel  and is calculated in Ref. \cite{Arbuzova:2021oqa} for different decay modes.  

Note, that this simple description of the decay by $\Gamma \dot R$-term in Eq. (\ref{ddot-R-Gam}) is valid only for harmonic oscillations of $R(t)$. For arbitrary  
law  of  evolution of $R$ 
the equation describing particle production by oscillating curvature becomes integro-differential, non-local in time 
{one}~\cite{Dolgov:1998wz,Arbuzova:2011fu}. 

Particle production leads to an appearence of the source term in Eq. \eqref{dot-rho} for the energy density:  
\be 
\dot\rho  + 3H (1+w) \rho  =  \bar S[R] \neq 0.
\label{dot-rho-pp}
\ee
The corresponding system of dimensionless equations takes the following form:
\be 
&& { { h' + 2h^2 =  - r/6, }}\label{h-prime-1} \\ 
&& {{ r'' + (3h + \gamma) r' + r = - 8 \pi \mu^2 (1-3w) y, }}\label{r-two-prime-1}\\  
&& {{y' + 3(1+w)h\,y = S[r], }}\label{y-prime-1}
\ee
where   ${ \mu = M_R/M_{Pl}}$, ${\gamma =  \Gamma/M_R}$. Detailed description of the solution of this system during the 
heating of the universe can be found in our review \cite{Arbuzova:2021etq}. 

Our analysis shows that the cosmological history in $R^2$-theory can be separated into 4 distinct epochs.   

Firstly, there was the inflationary stage, when the universe was void and dark with slowly 
decreasing curvature scalar, ${R(t)}$. The initial value of $ R$ should be rather large to ensure sufficiently long inflation: 
${R/M_R^2 \gtrsim 10^2}$.

The second epoch began when ${R(t)}$ approached zero and started to oscillate around it periodically changing sign:
\be 
R=-\frac{4M_R\cos(M_Rt+ \theta)}{t}, \, \ \ \  M_R = 3 \times 10^{13} \ \text{GeV}.
\label{Rsol}
\ee 
At this stage the Hubble parameter also oscillates almost touching zero: 
\be 
H= \frac{2}{3t} \left[1+ \sin(M_Rt + \theta )\right].  \label{Hsol}
\ee

The curvature oscillations resulted in the onset of creation of usual matter, 
that at this stage makes a subdominant contribution to the total cosmological energy density. 
We call this period the {\it scalaron dominated regime} at which the heating of the universe took place.  
During this time the universe evolution  and particle kinetics were
quite different from  those in general relativity. The new features of this stage
 open the window for heavy supersymmetry-kind particles to be the cosmological dark matter, modify high temperature baryogenesis, 
 lead to reconsideration of primordial black holes formation, {\it etc.}  

This period was followed by the transition from the scalaron domination to the dominance of the produced matter of 
mostly relativistic particles.
The oscillations of all relevant quantities  damped down exponentially
and the particle production by curvature   became negligible.

Lastly, after complete decay of the scalaron we arrived to the conventional cosmology governed by  general relativity. 

In the following sections we consider the epoch of { the universe heating}.  We calculate the freezing of the massive species $ X$ in plasma, 
which was created by the scalaron decays into heavier particles, and find 
the bounds on the masses of $X$-particles, allowing them to form the cosmological dark matter, for several different decay modes of the scalaron.

\section{Kinetics and freezing of massive supersymmetry-kind relics in cosmic plasma and bounds on masses of DM particles \label{kinetic}}

In this section we consider the freezing of massive supersymmetry-kind relics in cosmic plasma and obtain the bounds on masses of DM particles, 
following our papers \cite{Arbuzova:2018apk,Arbuzova:2021etq,Arbuzova:2020etv}.

The energy density of the produced particles depends upon the dominant decay mode of the scalaron. If scalaron decays into 2 massless scalars minimally coupled to gravity, the decay width of the scalaron and the energy density of the produced scalars are correspondingly equal to: 
\be 
\Gamma_s = \frac{M_R^3}{24M_{Pl}^2}, \ \ \ \   \rho_{s} = \frac{M_R^3}{240 \pi t}.
\label{Gamma-s}
\ee
In the case of scalaron decay into a pair of fermions  with mass ${m_f}$ we have:
\be 
\Gamma_f = \frac{ M_R m_f^2 }{6 M_{Pl}^2 }, \ \ \ \ \rho_f = \frac{ M_R m_f^2}{120 \pi t}.
\label{Gamma-f}
\ee
If the scalaron decay is induced by the conformal anomaly then the decay width and the energy density of the produced massless gauge 
bosons are given by the expressions: 
\be
\Gamma_{an} = \frac{\beta_1^2 \alpha^2 N}{96\pi^2}\,\frac{M_R^3}{M_{Pl}^2} , \ \ \ \ 
\rho_{an} = \frac{\beta^2_1 \alpha^2 N}{4 \pi^2} \,\frac{M_R^3}{120 \pi t}, 
\label{Gamma-an}
\ee
where
$\beta_1$ is the first coefficient of the beta-function, $N$ is the rank of the gauge group, 
$ \alpha$ is the gauge coupling constant, which at high energies depends upon the theory.

The presented laws demonstrate much slower decrease of the energy density of matter than normally for relativistic matter at scalaron 
dominated regime, where $\rho \sim 1/a^4(t) \sim 1/t^{8/3}$, since the scale factor at SD stage $a(t) \sim t^{2/3}$.
It is ensured by  the influx of energy from the scalaron decay. 

Expressions \eqref{Gamma-s} - \eqref{Gamma-an} can be compared with the energy density of matter in the 
standard GR cosmology: 
\be
\rho_{GR} = \frac{3H^2M_{Pl}^2}{8 \pi} = \frac{3 M_{Pl}^2}{ 32 \pi t^2}.
 \label{rho-GR}
\ee

It is interesting to compare equations connecting temperature with time in general relativity and in $R^2$-cosmology for different 
expressions for energy density of matter. 

Equating critical energy density of matter in GR (\ref{rho-GR}), as well as energy densities of matter (\ref{Gamma-s}), (\ref{Gamma-f}), and (\ref{Gamma-an})
in $R^2$-theory, to the energy density of relativistic plasma with temperature $T$ in thermal equilibrium:
\be 
\rho_{therm} = \frac{\pi^2 g_* T^4}{30},
\label{rho-pl}
\ee
where $g_*$ is the number of relativistic species in the plasma,  $g_* \sim 100$, 
 we obtain  that the connection of the temperature with time has 
very different forms  in general relativity and in $R^2$-cosmology:
\be \label{tT-GR}
(t T^2 )_{GR} &=& \left(\frac{ 90}{32 \pi^3 g_*}\right)^{1/2} \,M_{Pl} = const; \\ 
(t T^4)_s &=& \frac{ M_R^3}{4 \pi^3 g_*} = const;\\
(t T^4)_f &=& \frac{M_R m_f^2}{4\pi^3 g_*} = const;\\ \label{tT-an} 
(t T^4)_{an} &=& \frac{0.78 }{\pi^2  g_*}\,  \alpha_R^2 M_R^3 = const.
\ee
We see that the canonical relation between the matter temperature and the cosmological time in general relativity, $T^2 t = C_{GR} $, 
is replaced in $R^2$-theory by the relation $T^4 t =  C_{R^2}$. Moreover, as it is followed from  (\ref{tT-GR}),  $C_{GR}$ is a universal constant, proportional to the Planck mass, while in $R^2$-cosmology the constant $C_{R^2}$ depends on the model and may be strongly different for the scalaron decay into non-conformal massless bosons, fermions or gauge bosons. Note, that in Eq.~(\ref{tT-an}) the coupling ${\alpha_R}$ is taken at the energies equal to the scalaron mass.

The freezing of massive species $X$ with  mass $M_X$ is governed by the following equation: 
\be  {
\dot n_X + 3H n_X = -\langle \sigma_{ann} v \rangle \left( n_X^2 - n^2_{eq} \right)}, \   
n_{eq} = g_s \left(\frac{M_X T}{2\pi}\right)^{3/2} e^{-M_X/T}, 
\label{dot-n-X}
\ee
where $n_X$ is the number density of $X$-particles, $n_{eq}$ is their equilibrium number density, $g_s$ is the number of spin states.
$ \langle \sigma_{ann} v \rangle$ is the thermally averaged annihilation cross-section of X-particles with $v$ being the center-of-mass velocity.

Equation (\ref{dot-n-X}) was derived in 1965 by Zeldovich \cite{Zeldovich:1965gev} and collaborators \cite{Zeldovic:1965rys,Zeldovic:1965UFN}. 
In 1977 it was applied to freezing of massive stable neutrinos \cite{Lee:1977ua,Vysotsky:1977pe} and after that this equation was named as 
the Lee-Weinberg equation, though justly it should be called the Zeldovich equation. 

If annihillation of non-relativistic particles proceeds in S-wave the thermal averaging over medium is not essential and the annihilation cross-section
can be estimated as:
\be 
\langle \sigma_{ann} v \rangle= \sigma_{ann} v = \frac{\alpha^2 \beta_{ann}}{M_X^2}. 
\label{sigma-ann-S}
\ee 
If annihilating particles are Majorana fermions and their annihilation proceeds in P-wave, 
the thermal averaged annihilation cross-section acquires the factor $T/M_X$:
\be 
\langle \sigma_{ann} v \rangle =  \frac{\pi \alpha^2 \beta_{ann}}{M_X^2} \,\frac{T}{M_X}. 
\label{sigma-ann-P}
\ee
Here ${\beta_{ann}}$ is a numerical parameter 
proportional to the number of annihilation channels, ${\beta_{ann} \sim 10}$. 

We assume that direct $X$-particle production by curvature is suppressed in comparison with inverse annihilation of light particles into 
$X\bar X$-pair and $X$ particles are produced as a result of secondary reactions in relatvistic plasma, created by the scalaron decays into 
heavier species. 

An important comment is of order here. There are two possible channels to produce massive stable $X$-particles: first, directly through the 
scalaron decay into a $X\bar X$-pair, and, second, by the inverse annihilation of relativistic particles in thermal plasma. 
Direct production of $X\bar X$-pair by scalaron leads to the conclusion that the energy density of $X$-particles in the present day universe would be equal to the 
observed energy density of dark matter 
\be
{\rho_X^{(0)} \approx \rho_{DM}}\approx 1 {\rm keV/cm}^3, 
\ee
if ${M_X \approx 10^7}$ GeV. On the other hand, for such a small mass thermal $X$-particle production (through inverse annihilation) would be too
strong and would result in very large density of $X$-particles. But for larger masses $\rho_X^{(0)}$ would be unacceptably larger than DM energy density. 

A possible way out from this "catch-22" is to find a mechanism to suppress the scalaron decay into a pair of $X$-particles. And such mechanism does exist.  
If $X$-particles are Majorana fermions, their direct production is forbidden, since  
oscillating curvature scalar creates particles only in a symmetric state, while Majorana fermions must be in an antisymmetric state. 

Firstly, let us consider the scalaron decay into massless non-conformal scalars (detailed calculations are presented in our paper \cite{Arbuzova:2018apk}) . The dimensionless Zeldovich equation (\ref{dot-n-X}) has the form:
\be
\frac{df}{dx} =  {  - \frac{0.12 g_s \alpha^2 \beta_{ann}}{\pi^3 g_*} \left(\frac{M_R}{M_X}\right)^3} \, \frac{f^2 - f_{eq}^2}{x^5},
\label{Zeq-s}
\ee
where $x=M_X/T$ is a dimensionless new variable and the new dimensionless function $f$ is introduced according to
\be
n_X = n_{in} \left(\frac{a_{in}}{a}\right)^3, 
\ee
where  $n_{in}$ is the value of $X$-particle density at $a=a_{in}$ and $T_{in}=M_X$, so the $X$-particles can be considered as relativistic and
thus 
\be
n_{in} = 0.12 g_s T^3_{in} = 0.12 g_s M_X^3. 
\label{n-in}
\ee

Taking the following values of parameters: $g_* =100$, $\alpha = 0.01$, $\beta_{ann} =10$, and $M_R = 3\times 10^{13}$~GeV, 
we estimate the present day energy density of $X$-particles as:
\be 
\rho_X  \approx   10^{10} \left(\frac{10^{10} {\rm {GeV} }}{M_X} \right)} \,{ \rm{GeV/cm^3}.
\label{rho-X}
\ee
 This value is to be compared with the observed energy density of dark matter, ${ \rho_{DM} \approx 1}$ keV/cm$^3$. We see that 
$X$-particles must have  huge 
mass much higher than the scalaron mass, 
$M_X \gg M_R$,  to make reasonable DM density. However, if ${M_X > M_R}$, then classical scalaron field can still create $X$-particles, but the probability of their production would be strongly suppressed and such LSP with the
mass somewhat  larger than ${M_R}$ could successfully make the cosmological dark matter.

As the next step, let us turn to the scalaron decay into fermions or massive conformal scalars (for details see \cite{Arbuzova:2018apk}). If  bosons are conformally invariant due to non-minimal coupling to curvature, as $\xi R \phi^2 $ with $\xi = 1/6$, they are not produced if their mass is zero. The probability of production of both bosons and fermions is proportional to their mass squared.  In what follows we confine ourselves to consideration of fermions only. The width and the energy density of the scalaron decay into a pair of fermions are given by  expressions (\ref{Gamma-f})  and the largest contribution into the cosmological energy density at scalaron dominated regime is presented by the  decay into the heaviest fermion species. 

We assume, that the mass of the LSP is considerably smaller than the masses of the other decay products, ${m_X < m_f}$, at least as 
${m_X \leq 0.1 m_f}$. In this case the direct production of $X$-particles by ${R(t)}$  can  be neglected and LSPs are dominantly produced by the secondary reactions in the plasma, which was  created by the scalaron decay into heavier particles.

Dimensionless kinetic equation for freezing of fermionic species takes the form:
\be 
\frac{df}{dx} =  {- \frac{\alpha ^2 \beta_{ann}}{\pi ^3 g_*} \, \frac{ n_{in}\,M_R m_f^2}{m_X^6} }\,  \, \frac{f^2 - f_{eq}^2}{x^5}, 
\label{df-dx-R2-2-f}
\ee 
where $n_{in}=0.09 g_s m_X^3$ is the initial number density of $X$-particles at  $T \sim m_X$. 

The contemporary energy density of $X$-particles can be approximately estimated as 
\be
\rho_X \approx 7 \times 10^{-9} \frac{m_f^3}{m_X M_R}\, {\rm cm^{-3}},
\label{rho-fr-fn} 
\ee
where we have taken $\alpha = 0.01$, $\beta_{ann} = 10$, $g_*=100$.

If in Eq.~(\ref{rho-fr-fn}) we put ${m_f = 10^5}$ GeV and ${m_X = 10^4}$~GeV, then  the energy density of $X$-particles will be much less than the 
observed dark matter energy density: $\rho_X \ll \rho_{DM}$. However, choosing ${m_X \sim 10^6}$ GeV, ${m_f \sim 10^7}$ GeV we obtain that 
${\rho_X}$ becomes comparable with 
the energy density of the cosmological DM, ${\rho_{DM}~\approx~1}$ keV/cm$^3$:
\be 
\rho_ X \approx 2.1\,  \left(\frac{m_f}{10^7\, {\rm GeV}}\right)^3 \left(\frac{10^6\, {\rm GeV}}{m_X}\right) \, {\rm \frac{keV}{cm^3}}.
\label{rho-fin}
\ee

In the case when the scalaron decay is induced by the conformal anomaly we use expressions~\eqref{Gamma-an} for the decay width and for the
energy density of the produced gauge bosons and solve  Zeldovich equation \eqref{dot-n-X}. We assume that 
$X$-particles are Majorana fermions, so the direct production of $X\bar X$-pair by scalaron is forbidden. 
$X\bar X$-pairs are produced through the inverse annihilation of relativistic particles in the
thermal plasma.
The result of calculations of the frozen number density of $X$-particles with mass $M_X$ in cosmic plasma, 
which was created by the scalaron decay into massless gauge bosons due to conformal anomaly, lead to the conclusion that 
$X$-particles may be viable candidates for the carriers of the cosmological dark matter if their mass is about ${M_X \sim 10^{11}}$~GeV.

As we have seen, the range of allowed masses of $X$-particles to form cosmological DM depends upon the dominant decay mode of  scalaron.
The results are  summarized in Table~\ref{table:table1}. 

\begin{table}[!h]
\begin{tabular}{|p{75mm}|p{70mm}|} 
\hline
\vspace{.1mm} { Dominant decay channel of the scalaron}\vspace{.2cm}  & \vspace{.1mm} {Allowed $M_X$ to form DM}\\  \hline
Minimally coupled scalars mode: 
\vspace{-2mm}  
\be \nonumber
\Gamma_{s} = \frac{M_R^3}{24 M_{Pl}^2} 
\ee
 \vspace{-.3cm}  
 & \vspace{3mm} $M_X \gtrsim M_R \approx 3\times 10^{13}$ GeV \vspace{.2cm} 
\\ 
\hline
Massive fermions mode: 
\vspace{-1mm}
\be \nonumber
\Gamma_f = \frac{m_f^2 M_R}{6 M_{Pl}^2}
\ee
 \vspace{-.3cm}  
 &  \vspace{3mm} $ M_X \sim 10^6$ GeV \\
 \hline
Gauge bosons mode: 
\vspace{-2mm}
\be \nonumber
\Gamma_{an} = \frac{\beta_1^2 \alpha^2 N}{96\pi^2}\,\frac{M_R^3}{M_{Pl}^2} 
\ee
\vspace{-.3cm}  &  \vspace{3mm} $M_X \sim 10^{11}$ GeV   \\
\hline
\end{tabular}
\caption{\label{table:table1} The range of masses of $X$-particles  allowed to form cosmological DM for different  dominant decay modes of the scalaron.}
\end{table}

According to our results, the mass of DM particles, with the interaction strength typical for supersymmetric ones, can
be in the range from $10^6$ to $10^{13} $ GeV. It is tempting to find if and how they could be observed. There are some possibilities to make $X$-particles visible. The first one may be connected with the annihilation effects in clusters of dark matter in galaxies and galactic halos,  in which, according to Refs.~\cite{Berezinsky:2014wya,Berezinsky:1996eg},  the density of DM is much higher than DM cosmological density. Another possibility is to consider 
the superheavy DM particles, which could have a lifetime long enough to manifest themselves as stable dark matter, but at the same time 
their decays could lead to  possibly observable contribution to the  UHECR spectrum. Furthermore, instability of superheavy DM particles can arise due to 
Zeldovich mechanism through virtual black holes (BH) formation~\cite{Zeldovich:1976vq,Zeldovich:1977be}. We investigate the latter possibility in the next section.

\section{Cosmic rays from heavy particle decays \label{CR-decays}}

We assume that  superheavy DM particles have been created by oscillating curvature scalar 
$R (t)$ in the model of the Starobinsky inflation~\cite{Starobinsky:1980te} with the action \eqref{action-R2}:
\be \nonumber
S (R^2) = -\frac{M_{Pl}^2}{16\pi} \int d^4 x \sqrt{-g}\,\left[R- \frac{R^2}{6M_R^2}\right].
\ee

We have seen in the previous section that for the dominant scalaron  decay
 into a pair of  fermions with mass $m_f \sim 10^7$~GeV, the dark matter  particles, produced in secondary reactions in plasma, 
 would have the necessary cosmological density if their mass is about $10^6$~GeV. The decay width of the scalaron in this case is given 
 by expression ~\eqref{Gamma-f}: 
  \be \nonumber
  \Gamma_f = \frac{m_f^2 M_R}{6 M_{Pl}^2}.
  \ee
 This result is obtained for fermions with masses much smaller than
 the scalaron mass.
 
 Now we are interested in the case when the scalaron decays create particles with mass about $10^{21}$ eV, that is the energy of 
 ultra high energy cosmic rays (UHECR). 
 Equation (\ref{Gamma-f}) for  the width of the scalaron decay into such superheavy 
leptons,  $L$,  with mass $M_L \sim M_R/2$ should be modified in the following way:
\be
 \Gamma_L = \frac{M_L^2 M_R}{6 M_{Pl}^2} \sqrt{1- \frac{4M_L^2}{M_R^2}}.
 \label{Gamma-L}
 \ee
 The phase space factor  $(1- 4M_L^2/M_R^2)^{1/2}$ makes it 
 possible to arrange the density of presumably DM particles $L$  equal to the observed density of dark matter.
 
 However, { with the canonical energy scale of gravitational interaction} with
$M_{Pl} = 1.22 \times 10^{19}$~GeV, { the life-time of such DM-particles turns out to be too long} to allow for
any observable  consequences of their decays. 

A possible way out could be opened by diminishing the 
fundamental gravity scale at small distances down to a lower value $M_* < M_{Pl}$. 
This could lead to a considerable increase of decay probability of DM-particles.

Usually dark matter particles are supposed to be {absolutely} stable. However, there exists a mechanism suggested by Ya.~B.~Zeldovich ~\cite{Zeldovich:1976vq,Zeldovich:1977be}, which 
leads to decay of any presumably stable particle through creation and evaporation of virtual black holes.  
However, the rate of the proton decay calculated in the canonical gravity, 
{with the energy scale equal to $M_{Pl}$,} is extremely small. 
It is shown in what follows that smaller scale of gravity and huge mass of DM particles {both}
lead to a strong amplification of the Zeldovich effect.

In our work \cite{Arbuzova:2023dif} we have shown that 
superheavy DM particles with  masses about  $10^{12}$ GeV may decay through the virtual black hole
 with the life-time only a few orders of magnitude longer than  the universe age. Decays of such particles could make essential contribution to UHECR.
It can be achieved in the theory, where  gravitational coupling goes up at small distances 
 or high energies.

We consider the model proposed in Refs. \cite{Arkani-Hamed:1998jmv,Antoniadis:1998ig}, where 
 the observable universe with the Standard Model particles  is confined to 
  a 4-dimensional brane embedded in a
(4+$d$)-dimensional bulk, while gravity  propagates throughout the
bulk.  In such scenarios, the Planck mass $M_{Pl}$ becomes an
effective long-distance 4-dimensional parameter and the relation with
the fundamental gravity scale $M_{\ast}$ is given by
\begin{eqnarray}
M_{Pl}^2\sim M_{\ast}^{2+d}R_*^d ,
\label{M-ast}
\end{eqnarray}
where $R_*$ is the size of the extra dimensions: 
\be
R_{*} \sim \frac{1}{M_*}\left(\frac{M_{Pl}}{M_*}\right)^{2/d}.
\label{R-of-M}
\ee
As we see below for future application we choose 
$M_* \approx 3\times10^{17}$ GeV, so $R_*\sim 10^{(4/d)}/M_* > 1/M_*$.

Angular fluctuations of the cosmic microwave background radiation (CMBR) imply the following value of the scalaron mass:
$M_R \approx 3\times 10^{13} $~GeV~\cite{Faulkner:2006ub}.  As it is shown  
{in the quoted paper \cite{Faulkner:2006ub},} 
the CMBR fluctuations are expressed through the Planck and the scalaron masses in the following way:
\be
\delta ^2 \sim \left(\frac{M_R}{M_{Pl}}\right)^2, 
\label{CMBR}
\ee
where $\delta $ is the amplitude of the scalar fluctuations fixed by the observations. Thus, in models where the fundamental gravitational coupling is 
determined by $M_*$, instead of $M_{Pl}$, the scalaron mass should be changed appropriately, 
$M_R^* =   3\times 10^{13}(M_*/M_{Pl})$~GeV. 

We are interested in the case when the scalaron decays create particles with energies $10^{21}$ eV, that is the energy of UHECR. Thus, the scalaron mass, $M_R^*$,  should be {at least} of the order $10^{12}$~GeV. To this end we need to choose 
\be {
M_* = M_{Pl}/30 \approx 3 \times 10^{17}\,\,\ {\rm GeV}. }
\label{M-star} 
\ee

Analogously to the proton decay, $p\rar l^+ \bar q q$, let us consider the folliwng decay of $X$-particle: $X\rar L^+ \bar q_* q_*$,  
described by the diagram presented in Fig. \ref{f:pr-decay-BH}. 
	\begin{figure}[h]
	\vspace{-7cm}
  	\centering
  		\begin{minipage}[b]{1\textwidth}
  		\includegraphics[width=\textwidth]{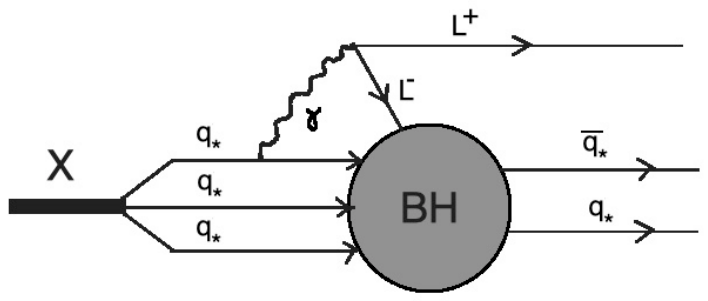}    
		 \end{minipage}
		 \vspace{-10cm}
		 \caption{{Diagram describing  $X$-particle decay into $L^+ \bar q_* q_*$ through virtual black hole.}}
		 \label{f:pr-decay-BH}
	\end{figure} 

{According to calculations of Ref. \cite{Bambi:2006mi}  the width of the proton decay
 into positively charged lepton and quark-antiquark pair is:} 
\be 
\Gamma(p\rar l^+ \bar q q) = 
\frac{m_p\,\alpha^2}{ 2^{12} \, \pi^{13}}
\left(\ln \frac{M_{Pl}^2}{m_q^2}\right)^2 \,
\left(\frac{\Lambda}{M_{Pl}}\right)^6 \,
\left(\frac{m_p}{M_{Pl}}\right)^{4+\frac{10}{d+1}}\, 
\int_0^{1/2} dx x^2 (1-2x)^{1+\frac{5}{d+1}},
\label{gamma-p}
\ee
 where $m_p \approx 1$GeV is the proton mass, $m_q \sim 300$ MeV is the constituent quark mass,  
 $\Lambda \sim 300$ MeV is the QCD scale parameter, $\alpha = 1/137$ is the fine structure constant,
 and $d$ is the number of "small' extra dimensions. {The QCD coupling constant $\alpha_s$ is supposed to be equal to unity.}
 {We can check that the proton decay rate is extremely small 
and the corresponding life-time is $7.3\times 10^{198} $ years which is
by far longer than the universe age, $t_U \approx 1.5\times 10^{10}$  years.}

 Using Eq. \eqref{gamma-p} {with the substitution $M_*$ instead of $M_{Pl}$} and keeping the same 
 values of other parameters, we can estimate the proton life-time with respect to decay 
 $p\rar \bar q q l^+ $ 
 { for $M_* \approx 3\times10^{17}$~GeV, $\tau = 2.17 \times 10^{188}$ years.}

 This case of decaying proton is mentioned for illustration only. We are interested in superheavy DM
 particles with masses about $10^{12}$ GeV and trying to formulate the scenario leading to their
 life-time with respect to the decay through the virtual BH only a few orders of magnitude longer than
 the universe age. 

{We consider the process  $X\rar L^+ \bar q_* q_*$ and assume that }
heavy dark matter $X$-particle with mass $M_X \sim10^{12}$ GeV consists of three heavy quarks, $q_*$, with comparable mass, 
{leaving $\Lambda_* $ as a free parameter.}
The life-time of $X$-particles can be evaluated using Eq.  \eqref{gamma-p} where we substitute 
$\alpha_*= 1/50$ instead of $\alpha = 1/137$, $M_X=10^{12}$ GeV  
instead of $m_p$, the mass of the constituent quark $m_{q_*} = 10^{12}$~GeV,
  and $d=7$: 
\be \label{tau-X}
\tau_X = \frac{1}{\Gamma_X} 
\approx {6.6\times 10^{-25} \rm{s} \, \cdot\frac{2^{10} \pi^{13}}{\alpha_*^2}} \left(\frac{\rm{GeV}}{M_X}\right)
\left(\frac{M_*}{\Lambda_*}\right)^6  \left(\frac{M_*}{M_X}\right)^{4+\frac{10}{d+1}}
\left( \ln \frac{M_*}{m_{q_*}} \right)^{-2} I(d)^{-1},
\ee
where we took {1/GeV = $6.6\times 10^{-25} $\rm{s}} and
\be
I (d) =\int_0^{1/2} dx x^2 (1-2x)^{1+\frac{5}{d+1}}, \,\,\, I(7) \approx 0.0057.
\label{i-of-d}
\ee
Now all the parameters, except for $\Lambda_*$, are fixed: $M_*=3\times 10^{17}$~GeV,
$M_X = 10^{12}$ GeV, $m_{q_*} \sim M_X$ and the life-time of X-particles can be estimated as: 
\be 
\tau_X \approx 7\times 10^{12}\,\,  {\rm years} \left(10^{15}\,\rm{GeV} /\Lambda_* \right)^6 \ \ \ vs \ \ \ 
t_U \approx 1.5\times 10^{10}\ {\rm years}. 
\label{tau-x-2}
\ee
A slight variation of $\Lambda$ near {$10^{15}$ GeV} allows to fix the life-time of the dark matter 
$X$-particles in the interesting range. They would be stable enough to behave as the cosmological dark 
matter and their decay could make considerable contribution into cosmic rays at ultra high energies. 

\section{Conclusions \label{concl}} 

\begin{itemize} 
\item
The existence of stable particles
with interaction strength typical for SUSY and heavier than several TeV is in tension with conventional Friedmann cosmology. 
\item
$R^2$-gravity
opens a way to save life of such $X$-particles, because in this theory
the density of heavy relics  with respect to the plasma entropy could be noticeably diluted by  radiation from the scalaron decay.
\item
The range of allowed masses of $X$-particles to form cosmological DM depends upon the dominant decay mode of  scalaron.
\item
In the model of high dimensional gravity modification there may exist superheavy
DM particles stable with respect to the conventional particle interactions. However, such DM
particles could decay though the virtual BH formation. 
\item
With a proper choice of the parameters 
the life-time of such quasi-stable particles may be larger than the universe age only by 3-4 orders of magnitude.
\item
This permits X-particles to make an essential contribution to the flux of ultra high energy cosmic rays.
\item
The considered mechanism may lead to efficient creation of
cosmic ray neutrinos of very high energies observed at IceCube
and Baikal detectors.
\end{itemize}

\section*{Acknowledgement}

This work was supported by the RSF Grant 22-12-00103.


\begin{thebibliography}{99}

\bibitem{Lin:2019uvt}
Lin, T. Dark matter models and direct detection. \textit{PoS} \textbf{2019}, \textit{333}, 009
[arXiv:1904.07915 [hep-ph]].

\bibitem{Fox:2019bgz}
Fox, P.J. TASI Lectures on WIMPs and Supersymmetry. Proceedings of the Theoretical Advanced Study Institute in Elementary Particle Physics: Theory in an Era of Data (TASI 2018), Colorado, 4-29 June 2018; {PoS(TASI2018)}: Trieste, Italy, 2019, 005.

\bibitem{Cline:2018fuq}
Cline, J.M. TASI Lectures on Early Universe Cosmology: Inflation, Baryogenesis and Dark Matter. Proceedings of the Theoretical Advanced Study Institute in Elementary Particle Physics: Theory in an Era of Data (TASI 2018), Colorado, 4-29 June 2018; {PoS(TASI2018)}: Trieste, Italy, 2019, 001
[arXiv:1807.08749 [hep-ph]].

\bibitem{Workman:2022ynf}
R.~L.~Workman \textit{et al.} [Particle Data Group],
``Review of Particle Physics,''
PTEP \textbf{2022} (2022), 083C01.

\bibitem{Arbuzova:2018ydn}
E.~V.~Arbuzova, A.~D.~Dolgov and R.~S.~Singh,
``Distortion of the standard cosmology in $R+R^2$ theory,''
\textit{JCAP} {\bf 07} (2018) 019
[arXiv:1803.01722 [gr-qc]].

\bibitem{Arbuzova:2018apk}
E.~V.~Arbuzova, A.~D.~Dolgov and R.~S.~Singh,
``Dark matter in $R+R^2$ cosmology,''
\textit{JCAP} {\bf 04} (2019) 014
[arXiv:1811.05399 [astro-ph.CO]].

\bibitem{Arbuzova:2020etv}
E.~V.~Arbuzova, A.~D.~Dolgov and R.~S.~Singh,
``Superheavy dark matter in $R+R^2$ cosmology with conformal anomaly,''
\textit{Eur. Phys. J. C} {\bf 80} (2020) no.11, 1047
[arXiv:2002.01931 [hep-ph]].

\bibitem{Arbuzova:2021etq}
E.~Arbuzova, A.~Dolgov and R.~Singh,
``$R^2$-Cosmology and New Windows for Superheavy Dark Matter,''
\textit{Symmetry} {\bf 13} (2021) no.5, 877.

\bibitem{Gurovich:1979xg}
V.~T.~Gurovich and A.~A.~Starobinsky,
``QUANTUM EFFECTS AND REGULAR COSMOLOGICAL MODELS,''
\textit{Sov. Phys. JETP} \textbf{50} (1979), 844-852. 

\bibitem{Starobinsky:1980te}
A.~A.~Starobinsky,
``A New Type of Isotropic Cosmological Models Without Singularity,''
\textit{Phys. Lett. B} {\bf 91} (1980) 99-102.

\bibitem{Faulkner:2006ub}
T. Faulkner, M. Tegmark, E.\,F. Bunn, and Y. Mao,
``Constraining $f(R)$ gravity as a scalar-tensor theory,"
\textit{Phys. Rev. D}\,{\bf 76} (2007) 063505
[arXiv:astro-ph/0612569].

\bibitem{Linde:2005ht}
A.D. Linde, 
`` Particle physics and inflationary cosmology. 
\textit{Contemp. Concepts Phys.} 5 (1990) 1-362
[arXiv:hep-th/0503203 [hep-th]].

\bibitem{Arbuzova:2021oqa}
E.~V.~Arbuzova, A.~D.~Dolgov and A.~S.~Rudenko,
``Calculations of Scalaron Decay Probabilities,''
Phys. Atom. Nucl. \textbf{86} (2023) no.3, 266-276
[arXiv:2112.11288 [hep-ph]].

\bibitem{Dolgov:1998wz}
A.~D. Dolgov, S.~H. Hansen, 
``Equation of motion of a classical scalar field with back reaction of produced particles", 
\textit{Nucl. Phys. B} {\bf 548} (1999), 408-426 
[arXiv:hep-ph/9810428 [hep-ph]].

\bibitem{Arbuzova:2011fu}
E.~V. Arbuzova, A.~D. Dolgov, L. Reverberi,  
``Cosmological evolution in $R^2$ gravity'',
\textit{JCAP} {\bf 02} (2012)  049
[arXiv:1112.4995 [gr-qc]].

\bibitem{Zeldovich:1965gev}
  Y.~B.~Zeldovich, ``Survey of Modern Cosmology'', \textit{Adv.\ Astron.\ Astrophys.} {\bf 3} (1965) 241.


\bibitem{Zeldovic:1965rys}
  Y.~B.~Zeldovich, L.~B.~Okun and S.~B.~Pikelner,
  ``Quarks, astrophysical and physico-chemical aspects,''
 {\it  Phys.\ Lett.\ }  {\bf 17} (1965) 164.

\bibitem{Zeldovic:1965UFN}
  Y.~B.~Zeldovich, L.~B.~Okun and S.~B.~Pikelner,
  ``Quarks, astrophysical and physico-chemical aspects,''
 {\it  Usp.\ Phys.\ Nauk. } {\bf 87} (1965) 113 [Sov.\ Phys.\ Usp. {\bf 8} (1965) 702. 

\bibitem{Lee:1977ua}
  B.~W.~Lee and S.~Weinberg,
  ``Cosmological Lower Bound on Heavy Neutrino Masses,''
 {\it  Phys.\ Rev.\ Lett.\ }  {\bf 39} (1977) 165.
  
\bibitem{Vysotsky:1977pe}
  M.~I.~Vysotsky, A.~D.~Dolgov and Y.~B.~Zeldovich,
  ``Cosmological Restriction on Neutral Lepton Masses,''
 {\it  JETP Lett.\ } {\bf 26} (1977) 188
   [{\it Pisma Zh.\ Eksp.\ Teor.\ Fiz.\  }{\bf 26} (1977) 200].
   
\bibitem{Berezinsky:1996eg}
V.~Berezinsky, A.~Bottino and G.~Mignola,
``On neutralino stars as microlensing objects,''
\textit{Phys. Lett. B} {\bf 391} (1997), 355-359.
doi:10.1016/S0370-2693(96)01495-5

\bibitem{Berezinsky:2014wya}
V.~S.~Berezinsky, V.~I.~Dokuchaev and Y.~N.~Eroshenko,
``Small-scale clumps of dark matter,''
\textit{Phys. Usp. } {\bf 57} (2014), 1-36
[arXiv:1405.2204 [astro-ph.HE]].

\bibitem{Zeldovich:1976vq}
Ya.B. Zeldovich, 
``A new type of radioactive decay: gravitational annihilation of baryons'', 
\textit{Phys. Lett. A} \textbf{59} (1976) 254. 

\bibitem{Zeldovich:1977be}
Ya.B. Zeldovich, 
``A Novel Type of Radioactive Decay: Gravitational Baryon Annihilation'', 
\textit{Zh. Eksp. Teor. Fiz.} \textbf{72}  (1977) 18.

\bibitem{Arbuzova:2023dif}
E.~V.~Arbuzova, A.~D.~Dolgov and A.~A.~Nikitenko,
``Cosmic rays from heavy particle decays,''
[arXiv:2305.03313 [hep-ph]].


\bibitem{Arkani-Hamed:1998jmv}
N.~Arkani-Hamed, S.~Dimopoulos and G.~R.~Dvali,
``The Hierarchy problem and new dimensions at a millimeter,''
Phys. Lett. B \textbf{429} (1998), 263-272
[arXiv:hep-ph/9803315 [hep-ph]].

\bibitem{Antoniadis:1998ig}
I.~Antoniadis, N.~Arkani-Hamed, S.~Dimopoulos and G.~R.~Dvali,
``New dimensions at a millimeter to a Fermi and superstrings at a TeV,''
Phys. Lett. B \textbf{436} (1998), 257-263
[arXiv:hep-ph/9804398 [hep-ph]].

\bibitem{Bambi:2006mi}
C.~Bambi, A.~D.~Dolgov and K.~Freese,
``A Black Hole Conjecture and Rare Decays in Theories with Low Scale Gravity,''
Nucl. Phys. B \textbf{763} (2007), 91-114
[arXiv:hep-ph/0606321 [hep-ph]].


\end{thebibliography}
\end{document}